\documentclass{elsart}

\usepackage{epsfig}

\usepackage{amssymb}
\usepackage{amsmath}

\begin{document}

\begin{frontmatter}



\title{Coalitions in the quantum Minority game: classical cheats and quantum bullies}


\author{Adrian P. Flitney\corauthref{cor}\thanksref{label2}}
\corauth[cor]{Corresponding author.}
\ead{aflitney@physics.unimelb.edu.au}
\ead[url]{http://aslitney.customer.netspace.net.au/default.html}
\author{Andrew D. Greentree\thanksref{label3}}

\address[label2]{School of Physics,
	University of Melbourne, Parkville, VIC 3010, Australia}
\address[label3]{Centre for Quantum Computer Technology and School of Physics,
	University of Melbourne}

\begin{abstract}
In a one-off Minority game,
when a group of players agree to collaborate
they gain an advantage over the remaining players.
We consider the advantage obtained in a quantum Minority game
by a coalition sharing an initially entangled state
versus that obtained by a coalition that uses classical communication
to arrive at an optimal group strategy.
In a model of the quantum Minority game where the final measurement basis is randomized,
quantum coalitions outperform classical ones when carried out by up to four players,
but an unrestricted amount of classical communication is better for
larger coalition sizes.
\end{abstract}

\end{frontmatter}

% main text
%\section{}
%\label{}
\section{Introduction}
\label{sec:intro}
Quantum game theory is the marriage of game theory,
the branch of mathematics dealing with conflict or competition scenarios,
with quantum mechanics.
Quantum game theory was initially applied to
classical game situations~\cite{meyer99,eisert99,marinatto00,kay01,iqbal01a,iqbal02a,flitney02a,du02a}
and to markets~\cite{piotr02a,piotr02b},
the traditional motivator of game theory~\cite{neumann44}.
In addition, it has been used to shed light on
the role of entanglement~\cite{han02a,du01b,du02b,du03b,flitney03a},
decoherence and noise~\cite{johnson01,chen02b,chen03a,ozdemir04a,flitney05},
and quantum correlations~\cite{shimamura04a,ozdemir04b,iqbal04a,iqbal04b},
among other phenomena~\cite{li01,iqbal02d,lee03a}.
An introduction to early work in quantum game theory is given in Ref.~\cite{flitney02c},
while a good description of the formalism of quantum games is provided by Lee and Johnson~\cite{lee03b}.

Originally introduced by Challet and Zhang in 1997~\cite{challet97},
the Minority Game (MG) has been extensively studied as a model of market behaviour~\cite{johnson98,savit99,challet00}.
In the MG the idea is that the majority is not always right,
in fact it is always wrong!
For example, in a stock market when most participants are buying,
the price of shares is forced up
and it is a good time to be a seller,
while when most agents are selling the price is depressed and it is a good time to be a buyer.
Minority-like games occur frequently in everyday life,
for example in choosing which of two routes to drive into the city
or which of two network printers to use for a print job.
Formally, in the MG a group of agents make simultaneous choices between two alternatives.
Those that select the minority alternative score one point
while those that select the majority score zero.
If the numbers are balanced, everyone is a loser.
In a one-off classical MG the best the players can do is to base their choice on the toss of a coin.
In a sequence of games
the strategies can be adaptive,
based on knowledge of previous selections,
and the agents are assumed to base their decision
on a certain fixed length history of game results.
Fluctuations around the optimal situation,
where just less than half of the players select the minority option,
can be minimized by adaptive strategic behaviour.
The literature on the classical MG is extensive:
for reviews see Challet {\it et al.}~\cite{challet00} or Moro~\cite{moro04}.

A four player quantum MG,
where the agents initially share a four-partite GHZ state,
was introduced by Benjamin and Hayden~\cite{benjamin01b},
and extended to $N$ players by Chen {\it et al.}~\cite{chen04}.
It was found that for an even number of agents,
new Nash equilibria (NE) appear that have no classical analogue
and that have payoffs superior to the best classical strategy.
The new NE even survive in the presence of decoherence~\cite{flitney06}.
However, for odd $N$ the quantum MG with an initial GHZ state offers
no improvement over the classical game.

The MG is one of the simplest games that is easily extendible to an arbitrary number of agents,
hence, apart from possible applications in quantum control,
one motivation for studying the quantum MG is to explore the features of multi-partite entanglement
in a game theoretic setting.
In the present work we study
strategies involving coalitions in the quantum game:
such coalitions can either be quantum,
where an entangled state is used as a resource;
or classical where the players are allowed to communicate their strategies via classical channels.
Through this means we hope to shed some light on the nature
of the advantage of entanglement over classical correlations.

\section{Quantum Minority Game}
\label{sec:qmg}
The standard quantum extension of the $N$-player Minority game~\cite{benjamin01b,chen04,flitney06}
is for the players to share a GHZ state $(|0\rangle^{\otimes N} + i |1\rangle^{\otimes N})/\sqrt{2}$
upon which each player acts with a single qubit unitary operator.
In the current work we consider a more general situation
where there can be arbitrary entanglement in the initial state
or where some players share an entangled set of qubits while the other qubits are separable.

Our model is described as follows.
An initial state is prepared with,
as usual,
one qubit for each player.
Each player can apply a local unitary operator,
representing their move, to their qubit.
A measurement of each qubit is then taken in the same orthonormal basis,
to be discussed later.
After the measurement,
the value of the $k$-th qubit is the ``choice'' made by the $k$-th player. 
Payoffs are then assigned in the same manner as in the classical MG,
with all players in the minority scoring one and the remaining players zero.
The state of the system is
\begin{equation}
|\psi_f\rangle = (\hat{M}_{1} \otimes \hat{M}_{2} \otimes \ldots \otimes \hat{M}_{N}) |\psi_i\rangle,
\end{equation}
where $|\psi_{i}\rangle$ and $|\psi_{f}\rangle$ are the initial and final states, respectively,
and $\hat{M}_{k}, \; k = 1, \dots, N$ is the move of the $k$-th player.
The expectation value of the $k$-th player's payoff is
\begin{equation}
\langle\$\rangle = \sum_{\xi_k} |\langle\psi_f|\xi_k\rangle|^2,
\end{equation}
where the summation is over all states $|\xi_{k}\rangle$ where the $k$-th player's selection
is in the minority.
A pure quantum strategy is described completely by the unitary operator selected by the player
to act on their qubit.
One way of parameterizing this operator is~\cite{flitney02c}
\begin{equation}
\label{e-qstrategy}
\hat{M}(\theta, \alpha, \beta) =
	\left( \begin{array}{cc}
		e^{i \alpha} \cos (\theta/2) & i e^{i \beta} \sin (\theta/2) \\
		i e^{-i \beta} \sin (\theta/2) & e^{-i \alpha} \cos(\theta/2)
	\end{array} \right),
\end{equation}
where $\theta \in [ 0,\pi ]$ and $\alpha, \beta \in [ -\pi,\pi ]$,
and the $k$-th player's move is $\hat{M}_{k} \equiv \hat{M}(\theta_k, \alpha_k, \beta_k)$.

In a four player quantum MG with the initial state $(|0\rangle^{\otimes 4} + i |1\rangle^{\otimes 4})/\sqrt{2}$,
Benjamin and Hayden showed that
there is an optimal strategy~\cite{benjamin01b}:
\begin{equation}
\label{eq:ne4}
\begin{split}
\hat{s}_{\rm \scriptscriptstyle NE}
			&= \frac{1}{\sqrt{2}} \cos (\frac{\pi}{16})(\hat{I} + i \hat{\sigma}_x)
		 	 \:-\: \frac{1}{\sqrt{2}} \sin (\frac{\pi}{16})(i \hat{\sigma}_y + i \hat{\sigma}_z) \\
			&= \hat{M}(\frac{\pi}{2}, \frac{-\pi}{16}, \frac{\pi}{16}).
\end{split}
\end{equation}
The strategy profile
$\{ \hat{s}_{\rm \scriptscriptstyle NE}, \hat{s}_{\rm \scriptscriptstyle NE},
\hat{s}_{\rm \scriptscriptstyle NE}, \hat{s}_{\rm \scriptscriptstyle NE} \}$
is a Nash equilibrium (NE),
a strategy profile from which no player can improve their result by a unilateral change of strategy~\cite{nash50}.
This NE gives an expected payoff of $\frac{1}{4}$ to each player,
the maximum possible from a symmetric strategy profile,
that is, one where all the players make the same choice.
The payoff in the quantum case is twice the maximum average payoff in the classical game,
where the best the players can do in a one-off game is selecting 0 or 1 at random.
The optimization results from the elimination
of the states
%from the final superposition
for which no player scores, 
that is, those where all the players make the same selection,
$|0\rangle^{\otimes 4}$ or $|1\rangle^{\otimes 4}$, 
or where the choices are evenly balanced,
$|0011\rangle, \: |0101\rangle$ etc.

This result can be extended to quantum Minority games with an arbitrary even number of players
that initially share a GHZ state~\cite{chen04}.
In our notation
the strategies
$\hat{M}(\pi/2, \eta-\delta, \eta+\delta)$,
where $\delta = (4 n + 1) \pi/(4 N), \; n = 0, \pm 1, \pm 2, \ldots$
and $\eta \in \{ -\pi+\delta, \pi-\delta \}$
are symmetric Nash equilibria.
The simplest and most obvious such strategy,
and one therefore that would be a focal point for the players' selection,
is $\hat{M}(\pi/2, -1/(4 N), 1/(4 N))$,
i.e., $n = \eta = 0$.
When all players select this strategy the initial GHZ state is transformed into
an equal superposition of states
with an even number of zeros and ones when $(N \, {\rm mod} \, 4)=0$,
or an odd number of zeros and ones when $(N \, {\rm mod} \, 4)=2$.
For example,
in the four and six player cases
\begin{equation}
\label{eq:example46}
\begin{split}
&\left[ \hat{M}(\frac{\pi}{2}, -\frac{\pi}{16}, \frac{\pi}{16})^{\otimes 4} \right]
	\, \frac{1}{\sqrt{2}} (|0\rangle^{\otimes 4} \,+\, i |1\rangle^{\otimes 4}) \\
&\quad = \frac{i + 1}{4} ( |1,3\rangle \,-\, |3,1\rangle), \\
&\left[ \hat{M}(\frac{\pi}{2}, -\frac{\pi}{24}, \frac{\pi}{24})^{\otimes 6} \right]
	\, \frac{1}{\sqrt{2}} (|0\rangle^{\otimes 6} \,+\, i |1\rangle^{\otimes 6}) \\
&\quad = \frac{ \sqrt{2}(\sqrt{3} - i)}{16}
				( |0\rangle^{\otimes 6} \,-\, |4,2\rangle \,+\, |2,4\rangle \,-\, |1\rangle^{\otimes 6} ),
\end{split}
\end{equation}
where $|m,n\rangle$ is the symmetric state with $m$ zeros and $n$ ones,
for example, $|1,3\rangle = |0111\rangle + |1011\rangle + |1101\rangle + |1110\rangle$.
An irrelevant global phase has been dropped from the last line of Eq.~(\ref{eq:example46}).
The resulting expected payoffs are not the maximum that can be achieved,
except in the case of $N=4$,
but are superior to anything that can be achieved by the uncoordinated actions of players
in a one-off classical MG.
For example,
the expectation value of the NE payoff in the six player quantum MG is $\frac{5}{16}$ to each player
compared to the maximum achievable expected payoff of $\frac{1}{3}$
that is obtained when two players select one alternative
and the remaining four select the other.
By contrast the classical six player MG yields an NE payoff of $\frac{3}{16}$.
A comparison of classical, quantum and Pareto optimal payoffs is given in Ref.~\cite{flitney06}.

Let us now consider other forms of entanglement in the initial state.
In the three player game the W state,
\begin{equation}
\label{eq:W3}
|W_3\rangle = \frac{1}{\sqrt{3}} (|001\rangle \,+\, |010\rangle \,+\, |100\rangle),
\end{equation}
is a state that yields the optimal expected payoff of $\frac{1}{3}$
and is fair to all players.
The $\bar{W}$ state gives an equivalent result.
For these initial states the players' optimal strategy is to apply the identity operator.

The same result can be achieved without entanglement by using the mixed state
\begin{equation}
\label{eq:mixed3}
\rho = \frac{1}{3} ( |001\rangle \langle 001| \,+\, |010\rangle \langle 010|
			\,+\, |100\rangle \langle 100|).
\end{equation}
To distinguish between these cases consider a modification to the game where
the final measurement is not necessarily taken in the computational basis $\{|0\rangle, |1\rangle\}$
but instead in an arbitrary orthonormal basis.
It suffices to write the logical $|0\rangle$ and $|1\rangle$ states as
\begin{equation}
\label{eq:logical}
\begin{split}
|0\rangle_{\rm L} &= \cos \phi \, |0\rangle + i \sin \phi \, |1\rangle, \\
|1\rangle_{\rm L} &= i \sin \phi \, |0\rangle + \cos \phi \, |1\rangle,
\end{split}
\end{equation}
for some $\phi \in [-\pi, \pi]$.
We shall refer to the quantum MG with a final measurement basis of the form of Eq.~(\ref{eq:logical})
as a randomized quantum MG.
The measurement basis is the same for all qubits.
For this game, a modification of Eq.~(\ref{eq:W3}):
\begin{equation}
\label{eq:W3star}
|W_{3}^{*}\rangle = \frac{1}{\sqrt{3}} (|001\rangle \:+\: e^{2 \pi i/3} |010\rangle
				\:+\: e^{-2 \pi i/3} |100\rangle),
\end{equation}
still yields the maximal expected payoff of $\frac{1}{3}$
in the three player game,
since in any basis the probability of all three qubits having the same value is zero.
In this sense Eq.~(\ref{eq:W3star}) is a three particle generalization
of the two particle singlet state
$(|01\rangle - |10\rangle)/\sqrt{2}$.
Clearly no unentangled state can have this property
since it is a global property of the ensemble of qubits.
By comparison, the mixed state of Eq.~(\ref{eq:mixed3}),
when measured in the randomized basis (\ref{eq:logical}),
yields an expected payoff of $\frac{17}{72} \simeq 0.236$ to each player.
One generalization of Eq.~(\ref{eq:W3star}) to $N$ qubits is
\begin{equation}
\label{eq:WNstar}
|W_{N}^{*}\rangle = \frac{1}{\sqrt{N}} \sum_{k=1}^{N} \exp(2 (k-1) \pi i/N)
	|0\rangle_{1} \ldots |0\rangle_{k-1} |1\rangle_{k} |0\rangle_{k+1} \ldots |0\rangle_{N}.
\end{equation}
The amplitudes of the individual states in the superposition are the $N$-th roots of unity.
For $N>3$, Eq.~(\ref{eq:WNstar}) is not the unique generalization of (\ref{eq:W3star}).
Any set of amplitudes in (\ref{eq:WNstar}) whose vectors in the complex plane form a closed polygon
(i.e., that sum to zero)
yields a state that has zero probability of all the qubits being equal when measured
in an arbitrary orthonormal basis.
For even $N$ one such state is a W state but with alternating plus and minus signs.
However, for $N > 4$ Eq.~(\ref{eq:WNstar}) or equivalent cannot be optimal for the players
since there are too few players in the minority.

In an attempt to optimize the initial entangled state for a randomized quantum MG,
consider a superposition of all states with a set number of ones and zeros,
where particular conditions are placed on the amplitudes.
That is, consider an $N$ qubit state of the form
\begin{equation}
\label{eq:general}
\begin{split}
|\psi \rangle &= \sum_{jk} c_{jk} |N-k,k \rangle, \\
\mbox{\rm with} \sum_{j} c_{jk} &= 0 \quad \forall \, k,
\end{split}
\end{equation}
subject to the usual normalization condition
\begin{equation}
\sum_{jk} |c_{jk}|^2 = 1.
\end{equation}
The sum in (\ref{eq:general}) is over the ${}^N C_k$ states $|N-k,k \rangle$ having $k$ ones and $N-k$ zeros.
This superposition has a zero probability of all qubits being equal
when measured in a basis of the form of Eq.~(\ref{eq:logical})
for arbitrary $\phi$.
That is, we have eliminated two of the states for which all players have zero payoff.
By making the number of ones and zeros as close as possible
we hope to maximize the size of the post-measurement minority
in a randomized quantum MG.
We shall consider the use of such states in section~\ref{s:arbitrary_coalition}.

\section{Comparison of quantum bullying and classical cheating}
\label{s:results}
\subsection{Quantum and classical coalitions}
\label{s:coalitions}
Instead of simply analyzing a game where all players start with equal resources,
a useful task is to examine multiplayer games
where one group of players forms a coalition for their mutual benefit.
A group of players can form a coalition
in a multiplayer quantum MG
in two fundamentally different ways.
A coalition can ``cheat'' by using classical communication
to arrange an optimal group selection.
This may involve a degree of trust among the partners
or a disinterested external party
that transmits the selected moves to the players.
For example,
a two player coalition would agree to choose different alternatives,
either selecting $|01\rangle$ or $|10\rangle$.
In a four player MG this would guarantee
that if there was a minority
it would be one of the coalition.
Interestingly, the players not in the coalition have no chance of being
in the minority regardless of how they play!
If the non-coalition players make a selection at random,
the optimal classical play in a one-off MG,
then there will be a minority 50\% of the time
and the expected payoff to the members of the coalition will be $\frac{1}{4}$.
Alternately,
the same result can be achieved by two quantum players sharing a known Bell state.
The state can always be transformed into
\begin{equation}
\label{eq:singlet}
(|01\rangle - |10\rangle)/\sqrt{2}
\end{equation}
by the action of local unitary operators,
that is, by moves allowable in the quantum game.
When the final measurement is taken in the standard basis
the expected payoff is the same as for the case of classical communication.
However, if we measure instead in the logical basis of Eq.~(\ref{eq:logical}) with a random value of $\phi$,
the probability that the pair of classical cheats are measured in the same logical state is
\begin{equation}
\frac{1}{2 \pi} \int_{0}^{2 \pi} 2 \cos^2 \phi \, \sin^2 \phi \: {\rm d}\phi = \frac{1}{4},
\end{equation}
with the remaining probability of $\frac{3}{4}$ that their choices are still orthogonal.
The expected payoff for the each of the classical cheats is consequently $\frac{3}{16}$.
For the quantum players sharing the singlet state Eq.~(\ref{eq:singlet})
it is well known that the qubits are orthogonal in any orthonormal basis,
thus ensuring that the quantum coalition partners still have an expected payoff of $\frac{1}{4}$
if the other players make random choices.

The sum of the payoffs for all four players is the same in the latter case
as it is for a classical MG when the players make a random choice,
however, all the payoff goes to the quantum coalition at the expense of the other two players.
For this reason we call the quantum coalition ``quantum bullies''
to distinguish them from a classical coalition of ``classical cheats.''

It is interesting to consider
a four player randomized quantum MG where there are two coalitions of two players.
In each case we average the final probabilities over all $\phi \in [0,2 \pi]$
before assigning payoffs.
Two quantum coalitions do poorly despite both choosing the ``optimal'' state, Eq.~(\ref{eq:singlet}),
since the number of zeros and ones is balanced in any measurement basis.
With one classical and one quantum coalition
there will be no minority with a probability of $\frac{3}{4}$
while in the remaining cases one of the quantum bullies win,
thus giving $\langle \$ \rangle = \frac{1}{8}$ for the players in the quantum coalition
and $\langle \$ \rangle = 0$ for the others.
In this case, by colluding the classical players do not improve their payoff,
but they do damage the payoff to the quantum coalition.
Finally, with two classical coalitions there is some probability of a minority
when measuring in the basis (\ref{eq:logical}),
resulting in $\langle \$ \rangle = \frac{5}{64} \simeq 0.078$ to each player.

\subsection{Quantum coalitions sharing a GHZ state}
For $N>4$ a pair of classical cheats have some probability of both being in the minority
even when their qubits are in the same logical state post-measurement,
so their payoff is reduced by slightly less than the factor of $\frac{3}{4}$ in these cases.
For larger coalition sizes the situation is more complicated.
As observed in Refs.~\cite{chen04,flitney06},
GHZ states of an odd number of qubits
do not confer any advantage to the players
in a multiplayer quantum MG.
% with all players initially sharing a GHZ state.
When the number of qubits is even,
local unitary operations can transform the GHZ state into
either a symmetric superposition of states with an even number of zeros and ones,
or a symmetric superposition of states with an odd number of zeros and ones.
When the coalition size $n$ is divisible by four,
the former is best,
while when $(n \, {\rm mod} \, 4) = 2$,
the latter is preferred.

Consider an initial state $(|0\rangle^{\otimes n} + i |1\rangle^{\otimes n})/\sqrt{2}$ for even $n$.
Each player applies the operator $\hat{M}(\pi/2, -\delta, \delta)$,
for some $\delta \in \mathbb{R}$ to be determined.
Where $(n \,{\rm mod}\, 4) = 0$,
the amplitude of states with an odd number of zeros and ones
is proportional to $\cos(n \delta) + \sin(n \delta)$,
giving a probability proportional to $1 + \sin(2 n \delta)$,
while the amplitude of states with an even number of zeros and ones
is proportional to $\cos(n \delta) - \sin(n \delta)$,
giving a probability proportional to $1 - \sin(2 n \delta)$.
Hence, when
\begin{equation}
\label{eq:odd}
\delta = \frac{(4 m + 1) \pi}{4 n}, \qquad m = 0, \pm 1, \pm 2, \ldots,
\end{equation}
the odd states are maximized and the even states vanish,
while when
\begin{equation}
\label{eq:even}
\delta = \frac{(4 m + 3) \pi}{4 n}, \qquad m = 0, \pm 1, \pm 2, \ldots,
\end{equation}
the reverse is true.
The roles of Eqs.~(\ref{eq:odd}) and (\ref{eq:even}) are interchanged when $(n \,{\rm mod}\, 4) = 2$.
The simplest strategy,
and therefore the focal point strategy for the players to choose,
is obtained by setting $m=0$.

For example,
four players each having one qubit from the state $(|0000\rangle + i |1111\rangle)/\sqrt{2}$
can transform the state into
\begin{equation}
\label{eq:4state}
\begin{split}
|\psi\rangle &= (1 + i)(-|0000\rangle + |0011\rangle + |0101\rangle + |0110\rangle \\
 & \quad + |1001\rangle + |1010\rangle + |1100\rangle - |1111\rangle)/4,
\end{split}
\end{equation}
by each applying $\hat{M}(\pi/2, -3 \pi/16, 3 \pi/16)$ to their qubit.
Having the selections of the four player coalition evenly distributed between $|0\rangle$ and $|1\rangle$
is optimal in an $N>4$ player MG.
For example, in the $N=5$ case
where there is only one player not in the coalition,
the coalition players score an average of $\frac{2}{5}$
while the fifth player is always in the majority and emerges empty handed.
The possibility of $|0000\rangle$ or $|1111\rangle$ in Eq.~(\ref{eq:4state})
reduces the expected payoff to the quantum coalition members to $\frac{3}{8}$
as indicated in table \ref{t:qresults}.

In the absence of entanglement,
but allowing classical communication,
a four player coalition can arrange to select $|0011\rangle$ or a permutation.
In a five player game this would produce the optimal expected payoff of $\frac{2}{5}$
to the coalition members in the absence of the randomizing of the measurement basis.
With the randomizing of the logical $|0\rangle$ and $|1\rangle$ by Eq.~(\ref{eq:logical})
the payoff to the members of the classical coalition is reduced fractionally to $\frac{51}{128} \simeq 0.398$.
%as indicated in table~\ref{t:cresults}.

With an odd number of players sharing a GHZ state
there is no symmetric strategy profile
that can optimize the players' payoffs.
An asymmetric strategy profile,
one where the players in the coalition play different strategies,
suffers from the disadvantage that it requires some classical communication
in addition to the entanglement
for it to be carried out reliably.
Table~\ref{t:qresults} gives the best a quantum coalition
starting with an even GHZ state can achieve.
%while table~\ref{t:cresults} summarize the results for a classical coalition utilizing
%an unlimited amount of classical communication to select the optimal combined state.

\begin{table}
\begin{center}
\begin{tabular}{|c|l|rrrrrr|}
\hline
n     & $\delta$           & N=5            & 6              & 7               & 8                & 9                & 10 \\
\hline
2     & $\frac{\pi}{8}$    & $\frac{1}{2}$  & $\frac{5}{16}$ & $\frac{1}{2}$   & $\frac{11}{32}$ & $\frac{1}{2}$    & $\frac{93}{256}$ \\
%&&&&&&& \\
4     & $\frac{3 \pi}{16}$ & $\frac{3}{8}$  & $\frac{1}{4}$  & $\frac{11}{32}$ & $\frac{15}{64}$ & $\frac{49}{128}$ & $\frac{67}{256}$ \\
%&&&&&&& \\
6     & $\frac{\pi}{24}$   &                & $\frac{5}{16}$ & $\frac{5}{16}$  & $\frac{3}{16}$  & $\frac{45}{128}$ & $\frac{35}{128}$ \\
%&&&&&&& \\
8     & $\frac{3 \pi}{32}$ &                &                &                 & $\frac{11}{32}$ & $\frac{11}{32}$  & $\frac{67}{256}$ \\
\hline
\end{tabular}
\end{center}
\caption{The optimal payoffs for the quantum coalition sharing an $n$-qubit GHZ state
in an $N$-player randomized quantum Minority game.
All the coalition members play $\hat{M}(\pi/2, -\delta, \delta)$ with $\delta$ given in the second column.}
\label{t:qresults} 
\end{table}

\subsection{Quantum coalitions sharing an arbitrary entangled state}
\label{s:arbitrary_coalition}
To optimize the size of the minority post-measurement
the quantum coalition would want to share an entangled state that
has close to a balanced number of zeros and ones
and that has no possibility of all the qubits being aligned after a measurement in the randomized basis.
Consider a state of the form of Eq.~(\ref{eq:general}) with (close to) balanced numbers of zeros and ones:
\begin{equation}
\label{eq:optimal}
|\psi^* \rangle = \sum_{j=0}^{c-1} e^{2 j \pi i/c} |N-m, N+m \rangle/\sqrt{c},
\end{equation}
where $c = {}^N C_m$
and $m = N/2-1$ for even $N$, or $N/2-\frac{1}{2}$ for odd $N$.
The summation is over the $c$ states with $N-m$ zeros and $N+m$ ones.

A comparison of the expected payoffs in an $N$ player randomized quantum MG
for a coalition of $n$ quantum bullies using Eq.~(\ref{eq:optimal})
and $n$ classical cheats selecting the optimal (unentangled) state by using classical communication
is shown in figure \ref{f:bully}.
Observe that for the two player coalition,
the quantum bullies easily outperform the classical cheats,
showing the superiority of the singlet state for this task,
as discussed in the $N=4$ case in section~{\ref{s:coalitions}.
However, as $n$ increases the advantage of using the entangled state of Eq.~(\ref{eq:optimal}) diminishes.
For $N>5$, Eq.~(\ref{eq:optimal}) is no longer optimal.
It is an open question whether there is a different form of entangled state
that would outperform the classical coalition.
The quantum bullies can, however, always simply choose the same state as the classical cheats.
We conclude that the benefit of entanglement in comparison to classical communication
in this game diminishes as the coalition size increases,
that is, as the classical resources
(e.g., communication or memory)
required for the classical cheats to coordinate their actions increases.

As coalition size increases the classical coalition partners must utilize increasing
amounts of communication,
or memory and communication.
If some model of the cost of classical communication and storage,
as well as the cost of creating and transmitting the entangled quantum state,
was factored into the calculations a different picture may emerge.
However, this is beyond the scope of this work.

\begin{figure}
\begin{center}
\includegraphics[width=12cm]{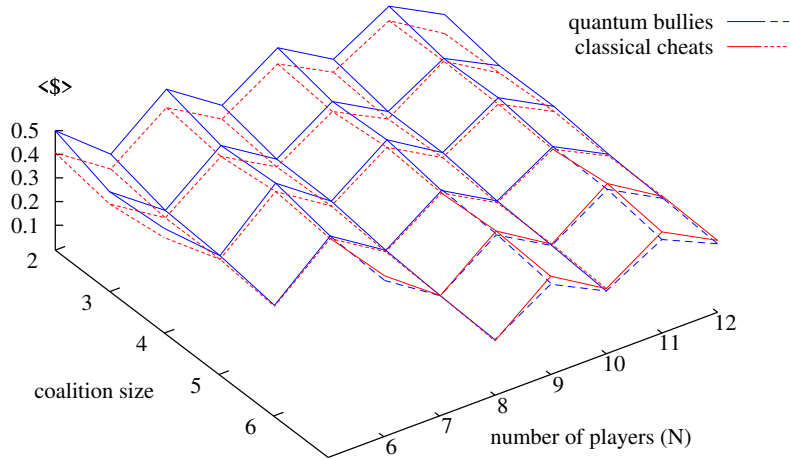}
\end{center}
\caption{\label{f:bully}The optimal payoff to each of the $n$ players in the coalition
in an $N$ player randomized quantum Minority game
when they share an entangled state (blue) of the form of Eq.~(\ref{eq:optimal})
or when the exploit classical communication (red)
to arrange an optimal state.
Dashed lines are hidden below solid lines.
The lines are only a guide for the eye since the payoffs only make sense for integer $n$ and $N$.}
\end{figure}

\section{Conclusion}
\label{sec:conc}
We have considered a modification of the quantum Minority game
where the final measurement is not taken in the computational basis
but rather in an arbitrary (randomized) orthonormal basis.
In such a scenario a comparison is made
of the benefit to a group of $n$ players colluding
by sharing an entangled set of $n$ qubits
as opposed to using classical communication to set up an optimal state.
A generalization of the two qubit singlet state to $n$ players
that eliminates the possibility of all the players ending up with the same choice
after the measurement
suggests itself as an optimal strategy.
For small coalition sizes the advantage of using entanglement is clear,
but as the coalition size increases
the generalized singlet state becomes non-optimal.
Indeed it is difficult to find an entangled state
that is more advantageous for the quantum coalition
than the simple unentangled choice made by a group of players colluding by classical communication.
However, this does not take into account the practicalities of such information sharing.
Accordingly, it is interesting to speculate on the results
where the cost of the operations
are taken into account.

The protocol of Eisert {\em et al.}~\cite{eisert99}
has been the dominant means by which quantum game theory has been explored.
In this paper we have moved beyond this
to explore different forms of entanglement and partial entanglement in the initial state.
We hope that the analysis of coalitions in a multi-player quantum game
can illuminate some of the advantages of entanglement over simple classical correlations
and shed some light on multi-partite entanglement.

\section*{Acknowledgments}
\noindent
Thanks go to
to our colleagues Austin Fowler and Lloyd Hollenberg
for their ideas and helpful discussion.
Funding for APF was provided by the Australian Research Council grant number DP0559273.
ADG was supported by the Australian Research Council, the Australian
Government and by the US National Security Agency (NSA), Advanced
Research and Development Activity (ARDA) and the Army Research
Office (ARO) under contract W911NF-04-1-0290.

% \appendix

% \section{}
% \label{}

\end{document}